\documentstyle[newarcrc,fleqn,psfig]{article}

\def\etal{{et\,al.}}
\def\msun{M$_{\odot}$}

\def\mdot{$\dot M$}

\def\degs{\ifmmode ^{\circ}\else$^{\circ}$\fi}
\newbox\grsign \setbox\grsign=\hbox{$>$}
\newdimen\grdimen \grdimen=\ht\grsign
\newbox\laxbox \newbox\gaxbox
\setbox\gaxbox=\hbox{\raise.5ex\hbox{$>$}\llap
     {\lower.5ex\hbox{$\sim$}}}\ht1=\grdimen\dp1=0pt
\setbox\laxbox=\hbox{\raise.5ex\hbox{$<$}\llap
     {\lower.5ex\hbox{$\sim$}}}\ht2=\grdimen\dp2=0pt
\def\gax{$\mathrel{\copy\gaxbox}$}
\def\lax{$\mathrel{\copy\laxbox}$}
\DeclareRobustCommand{\ion}[2]{%
\relax\ifmmode
\ifx\testbx\f@series
{\mathbf{#1\,\mathsc{#2}}}\else
{\mathrm{#1\,\mathsc{#2}}}\fi
\else\textup{#1\,{\mdseries\textsc{#2}}}%
\fi}
\def\arcsec{\ifmmode ^{\prime\prime}\else$^{\prime\prime}$\fi}
\def\v7{V751\,Cyg}
\def\rxj0513{RX\,J0513.9--6951}
\def\hd{HD 104994}

\title{V751 Cyg and V Sge as transient supersoft X-ray sources} 

\author{Jochen Greiner\address{Astrophysikalisches Institut Potsdam, \\
        14482 Potsdam, An der Sternwarte 16, Germany}
        }

\begin{document}
\maketitle

\begin{abstract}
I review the observational evidence for luminous, soft X-ray emission
during optical low-states in the two cataclysmic variables V751 Cyg and V Sge,
and discuss the possible link to the canonical supersoft X-ray sources.
\end{abstract}

\section{Introduction}

Supersoft X-ray sources (SSS) were established as a new class of astronomical
objects during the early years of this decade (Tr\"umper \etal\ 1991,
Greiner \etal\ 1991, Kahabka \& van den Heuvel 1997) and are thought
to contain white dwarfs accreting mass at
rates high enough to allow quasi-steady nuclear burning of the accreted
matter (van den Heuvel \etal\ 1992). The sources are highly luminous 
($L_{bol} \sim 10^{36}-10^{38}$ ergs s$^{-1}$), but since their
 characteristic temperatures are on the order of tens of eV,
much of the energy is radiated in
the far ultraviolet or soft X-ray region of the spectrum, where the
radiation is easily absorbed by the interstellar medium. Because of this, only
2 close-binary Galactic supersoft sources are known (Motch \etal\ 1994, 
Beuermann \etal\ 1995), though there should be about 
1000 in the Milky Way (Di\thinspace Stefano \& Rappaport 1994).
The situation is further complicated by the fact that supersoft X-ray binaries 
are highly time variable, both at X-ray and optical wavelengths (Greiner 1995).
The greatest number are known in the Magellanic Clouds, but they
are difficult to study because of their distance.
The quest to find new SSBs has inspired several projects
including comprehensive studies of deep ROSAT pointings. 
So far, these were not generally successful, however.

Recently, two other approaches to search for further members of the
SSS class have been attempted: 
(1) Using a unique variability pattern:
Although most SSS are variable in their X-ray and/or optical emission,
the behaviour of several systems is distinctive.
Data collected during the MACHO team's monitoring of the LMC has shown that
the seemingly sporadic X-ray bright states of RX J0513.9--6951, a well-known
supersoft X-ray binary in the LMC (Schaeidt \etal\ 1993,
Pakull \etal\ 1993), are correlated with short-lived optical low
states (Pakull \etal\ 1993, Reinsch \etal\ 1996, Southwell \etal\ 1996). 
As first suggested by Brian Warner, the birthday of whom we are celebrating
here, at the SSS workshop in Garching in 1996,
the 3 year MACHO light curve (Southwell \etal\ 1996) suggests a strong 
similarity to the  VY Scl stars.
VY Scl stars are a subclass of nova-like, cataclysmic variables which are
bright most of the time, but  occasionally drop in brightness by several
magnitudes  at irregular intervals (Bond 1980, Warner 1995).
(2) Searching among unusual cataclysmic variables: 
One possible Galactic source, V Sagitae, has been suggested
by studying the properties of several unusual cataclysmic variables 
(Patterson \etal\ 1998).
Steiner \& Diaz (1998) note the similarity of $3$ other Galactic systems
to V Sge. 

In the following, I will describe both strategies and review the evidence
for transient, luminous supersoft X-ray emission in the two 
cataclysmic variables (CVs) V751 Cyg and V Sge during their optical low states.

\section{V751 Cyg}

\begin{figure}[th]
    \vspace{-0.35cm}
    \centering{
    \hspace*{0.01cm}
    \vbox{\psfig{figure=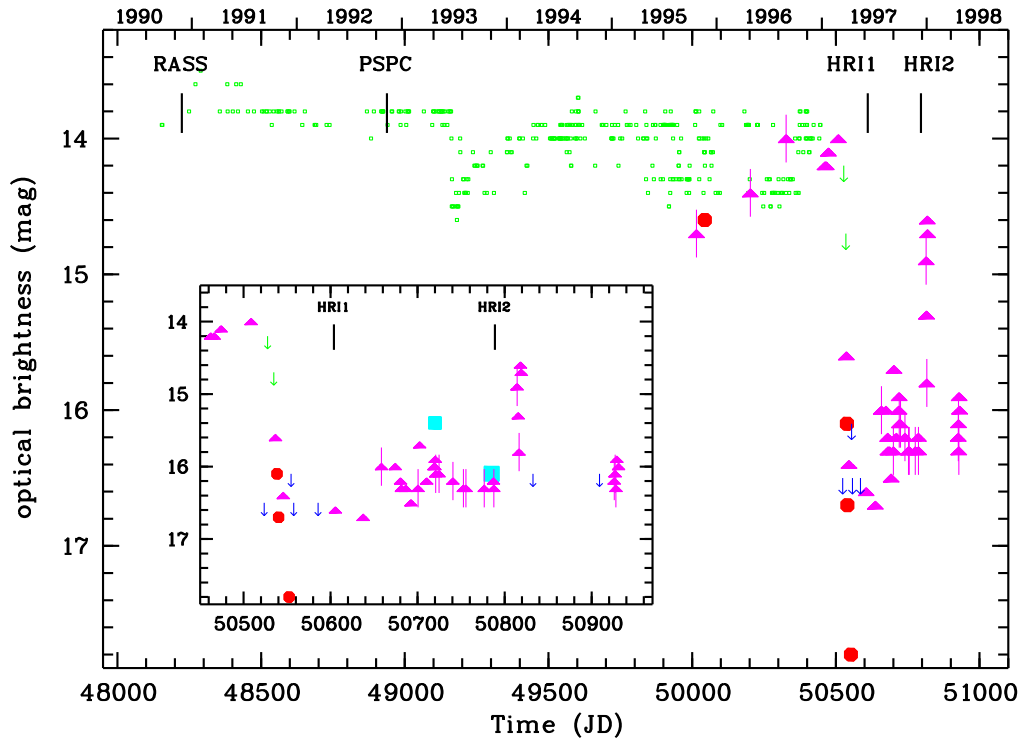,width=13.5cm,%
          bbllx=2.cm,bblly=1.9cm,bburx=17.7cm,bbury=13.5cm,clip=}}\par}
    \vspace*{-0.9cm}
    \caption[pha]{Optical light curve of V751 Cyg over 8 years:
         small (gray) squares denote measurements as reported to AFOEV 
        and triangles denote measurements as reported to VSNET.
       Large filled circles
        are CCD measurements of the Ouda team (also taken from VSNET).
        Arrows  denote  upper limits.
         At the top the times of the  ROSAT observations are marked.
         V751 Cyg is detected only in the HRI observations during the optical
        low state. The inset shows a blow-up of the optical low-state together
       with the times of the two HRI observations. The two squares in the inset
        represent the mean brightness on Sep. 29/30, 1997 as derived from
        spectra and Dec. 3, 1997 photometry (at the time of the second 
       HRI observation) (from Greiner \etal\ 1999).
        }
      \label{lc751}
   \end{figure}

Full details of the correlated X-ray and optical observations of \v7\ 
have appeared elsewhere (Greiner \etal\ 1999), 
so we only summarize the relevant information here.
The distinct lightcurve of \rxj0513\ and its similarity
to VY Scl stars led us to decide to monitor 
the light curves of the known VY Scl stars.
When \v7\ started to drop in brightness somewhere between
1 March  and 11 March 1997 (Fig. \ref{lc751})
we performed a target-of-opportunity ROSAT HRI observation 
(4660 sec) on 3 June 1997. A new X-ray source, RX J2052.2+4419, 
was discovered within 1\arcsec\ of V751 Cyg, at a mean count rate of
0.015 cts/s. During a second ROSAT HRI observation on Dec. 2--8, 1997 the
count rate and X-ray spectrum are nearly identical to the June values.

In contrast, \v7\ was not detected during the ROSAT all-sky survey
on Nov. 19/20, 1990 giving a 3$\sigma$ upper limit of 0.019 cts/s in the PSPC.
In addition, it was also not detected during a serendipituous pointing
on Nov. 11, 1992 providing an upper limit of 0.0058 cts/s in the PSPC.
On both occasions \v7\ was in its optical bright state. This suggests
an  anti-correlation of optical and X-ray intensity in \v7.

\begin{figure}
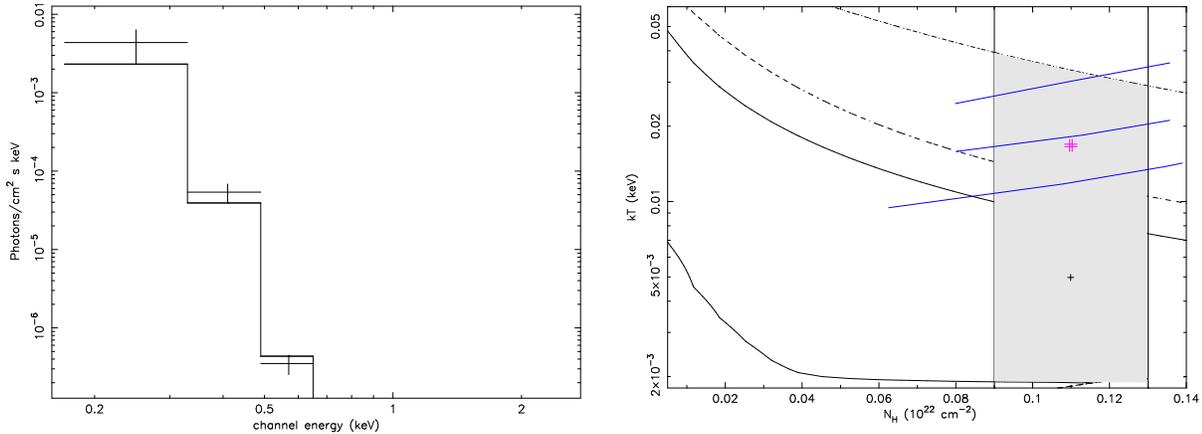

    \vbox{\psfig{figure=v751cyg_ufs.ps,width=0.49\textwidth,%
        bbllx=2.4cm,bblly=2.4cm,bburx=18.cm,bbury=13.8cm,clip=}}\par
    \vspace*{-5.75cm}\hspace*{0.5\textwidth}
    \vbox{\psfig{figure=v751cyg_cont.ps,width=0.49\textwidth,%
        bbllx=0.9cm,bblly=1.4cm,bburx=16.8cm,bbury=12.8cm,clip=}}\par
    \vspace*{-0.8cm}
    \caption[pha]{Spectral fit result of the June 1997 ROSAT observation
         of V751 Cyg:
       {\bf Left:} The photon spectrum deconvolved with a blackbody model with 
         $N_{\rm H}$ fixed at the value as derived from the IUE spectrum.
         {\bf Right:}
        Confidence contours of the blackbody fit in the
         $kT-N_{\rm H}$ plane: 68\% (solid line),
         90\% (long-dashed line), 99\% (short-dashed line). 
         The two vertical lines denote the $N_{\rm H}$ range allowed
         by the IUE data ((1.1$\pm$0.2)$\times$10$^{21}$ cm$^{-2}$), 
        and the vertical extent of the hatched region marks the 99\% confidence
         region of the blackbody temperature.
         The three solid  lines crossing the hatched region
        mark contours of constant luminosity (at an assumed distance of 500 pc)
         of 10$^{34}$, 10$^{36}$ and 10$^{38}$ erg/s (top to bottom).
       The double cross denotes the parameter pair used in the text.
         }
      \label{ufs}
      \vspace{-0.4cm}
   \end{figure}

A new method (Prestwich \etal\ 1999) to extract reliable spectral  
information from HRI data allowed to craft  a response matrix 
for a given observation.
Fits using this response matrix to all the source photons 
of V751 Cyg show that simple black-body 
models with kT of a few tens of eV are consistent with the data, 
whereas higher temperature models (0.5 keV) can be ruled out (Fig. \ref{ufs}).
An IUE observation performed in 1985 (during an optical high state)
was used to derive the extinction towards \v7\ based on
the broad absorption centered at  2200 \AA: $E(B-V)=0.25\pm0.05$
$\equiv$ $N_{\rm H}$ = 1.1$\times$10$^{21}$ cm$^{-2}$.
This implies a distance of the order of 500 pc.
With this $N_{\rm H}$ the X-ray spectral fitting gives
$kT = 15^{+15}_{-10}$ eV (see Fig. \ref{ufs}).
At this temperature, the bolometric luminosity on 3 June 1997 is 
6.5$\times$10$^{36}$ (D/500 pc)$^2$ erg/s. 

Thus, during its optical low state,
\v7\ was emitting soft X-rays with a temperature and luminosity
which confirm that it is a transient supersoft X-ray source.
The appearance of \ion{He}{ii} 4686 \AA\ emission in optical spectra taken
in Sep. 1997 also indicates the presence of $>$54 eV photons.
V751 Cyg, like the other members of the VY Scl star group,
accretes at a few times 10$^{-8} M_\odot$ yr$^{-1}$. If the mass of the
white dwarf in \v7\ is small, this may allow nuclear burning, as the
high X-ray luminosity suggests. 
The \v7\ values of $M_{\rm V}^{\rm max} = 3.9$  and
${\log \Sigma} = (L_x/L_{\rm Edd})^{1/2} P_{\rm orb}^{2/3} (hr) = -0.23$ are
consistent, within the uncertainties of $L_x$ and $P_{\rm orb},$ with the
relation $M_{\rm V} = 0.83(\pm 0.25) - 3.46 (\pm 0.56) \log \Sigma$
found for 5 SSB (van Teeseling \etal\ 1997) implying
that, if nuclear burning is the correct interpretation
of the X-ray flux during the optical low state,
then nuclear burning may continue during the optical high state.

The discovery that \v7\ is a transient supersoft X-ray source arose from the
similarity in the optical light curve of RX J0513.9--6951 and VY Scl stars.
\rxj0513 (Schaeidt \etal\ 1993, Pakull \etal\ 1993) shows $\sim$4 week 
optical low states which are accompanied by luminous supersoft X-ray 
emission (Reinsch \etal\ 1996, Southwell \etal\ 1996). It is generally 
assumed that the
white dwarf accretes at a rate slightly higher than the burning rate,
and thus is in an inflated state during the optical high state.
Changes in the irradiation of the disk caused by the expanding/contracting
envelope around the white dwarf have been proposed as explanation for the
1 mag intensity variation  (Reinsch \etal\ 1996, Southwell \etal\ 1996). 
The explanation of the X-ray/optical variability of \v7\ could be
similar to RX J0513.9--6951:
\mdot variations change both the photospheric radius and
the disk spectrum.
If the white dwarf has a small mass, than photospheric radius expansion
is reached  at 1$\times$10$^{-7}$ \msun/yr (Cassisi \etal\ 1998).

The explanation for the character of the optical and UV observations 
is not yet clear, but it seems certain that the illumination of the donor and 
disk play important roles in determining what we see.  
If the X-ray source during the optical low state indeed is very luminous
one may expect a strong heating effect on the secondary as well as on
the accretion disk.
The heating of the secondary in \v7\ is probably comparable to
that in SSS because the illumination depends on the
ratio of companion radius and binary separation which is similar in both
kind of systems. 
 Unfortunately, no photometry has been
obtained during the optical low state to immediately test for this effect 
in \v7\, though it is anyway not expected to produce a strong modulation 
due to the low orbital inclination.

The question of the illumination of the accretion disk has to be addressed
separately for optical low and high state.
There is ample evidence in some VY Scl stars that during the
optical low state the accretion disk has vanished. Though we have no direct
evidence for this in \v7\ due to the lack of optical observations, 
the disk is certainly optically thin thus
drastically reducing the efficancy of illumination.
In the optical high state the illumination depends on 
whether hydrogen burning stops or whether it continues on 
an inflated white dwarf at a temperature below the sensitivity range of ROSAT: 
If the  burning stops then there are no soft X-rays which 
could be reprocessed. If the nuclear burning continues, 
reprocessing may still not be strong because the amount of reprocessing 
depends on the size of the accretion disk (orbital period) and only for large
disks also on the flaring of the disk. For V751 Cyg the disk is so small
that any flaring is probably insignificant compared to the angle which the 
white dwarf subtends with respect to the disk. Even if flaring were
significant, reprocessing of the radiation from the
white dwarf will begin to have a dominant effect on the local disk
temperature (King 1997, Knigge \& Livio  1998) if the white dwarf luminosity
$L_{\rm WD}$ \gax $2.5 L_{\rm acc} (1-\beta)^{-1}$ (where $\beta$ is the albedo
of the disk surface). That is, a 
disk around a 1 \msun\ white dwarf accreting at 10$^{-8}$ \msun/yr 
will be dominated by reprocessing only if the white dwarf temperature is 
$>$2$\times$10$^5$ K $\equiv$ 17 eV.
This is seemingly just a value between the 
temperatures of SSS (30--50 eV) and V751 Cyg (15 eV). 
Thus, one difference of the systems could be that the disk in V751 Cyg
is not flared and therefore not 
dominated by reprocessing while the SSS disks are flared and
dominated by reprocessing and thus are optically much brighter than 
the VY Scl disks.

\section{V Sge}

It has recently been suggested (Steiner \& Diaz 1998; Patterson \etal\ 1998)
that V Sge (and possibly also WX Cen, V617 Sge and \hd) have
properties very similar to SSBs.
These suggestions are based on the following characteristics,
shared by these four stars but rare or even absent among canonical
cataclysmic variables:
(1) the presence of both \ion{O}{VI} and \ion{N}{V} emission lines,
(2) a \ion{He}{II} $\lambda 4686$/H$\beta$ emission line ratio $>$2,
(3) rather high absolute magnitudes and very blue colours, and
(4) orbital lightcurves which are characterized by a wide and deep eclipse.

   \begin{figure}[thb]
   \centering{
   \hspace{0.01cm}
    \vbox{\psfig{figure=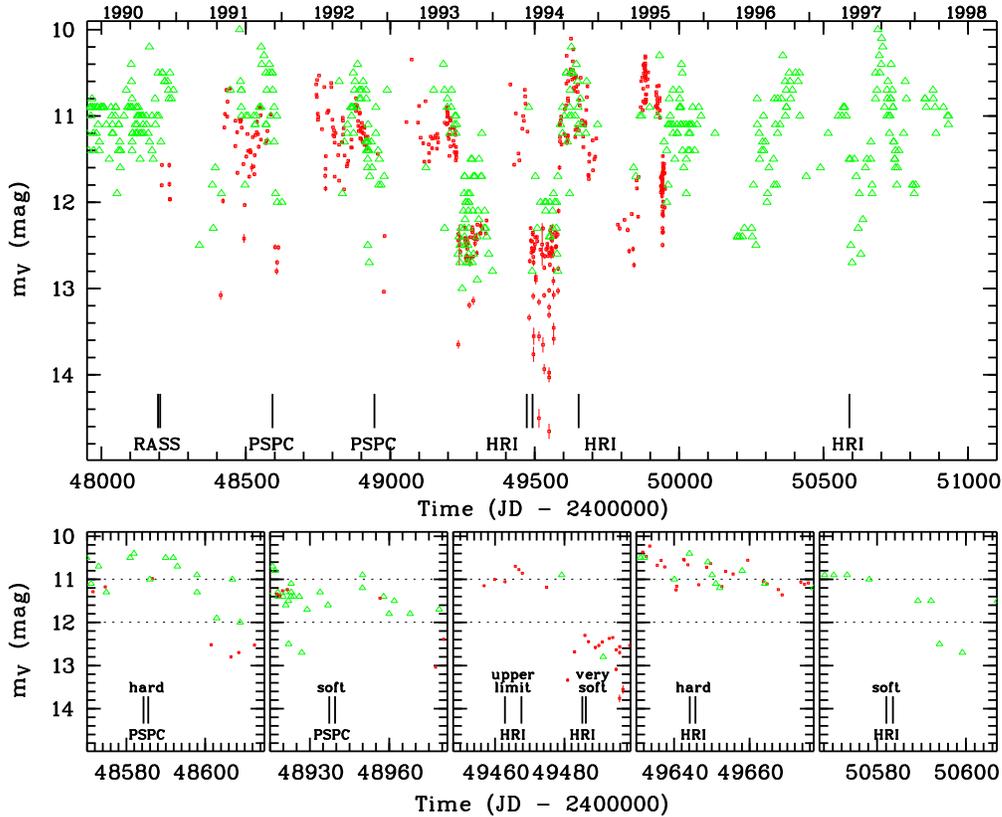,width=0.83\textwidth,%
           bbllx=1.1cm,bblly=1.5cm,bburx=19.1cm,bbury=16.4cm,clip=}}\par}
    \vspace*{-0.9cm}
    \caption[lcvsge]{Optical light curve of V~Sge with data from
            Robertson (\etal\ 1997 and unpublished; red dots) and 
            VSOLJ (Web; green triangles).
            Vertical dashes mark the times of ROSAT observations.
            The lower panels show blow-ups around the ROSAT observations
            (two vertical
            lines mark the start and the end of each ROSAT exposure).
            The dotted lines denote the boundaries of the three optical states.
            (From Greiner \& Teeseling 1998)
           }
      \label{lcvsge}
   \end{figure}

V~Sge has been the target of three dedicated pointed PSPC and HRI observations
(one of these splits into 3 separate observation intervals),
and in addition is in the field of view of another PSPC observation.
A detailed comparison of the optical states of V Sge and archival ROSAT
observations has shown that during optical bright states, V~Sge is a faint 
hard X-ray source, while during optical faint states ($V>12$ mag), 
V~Sge is a `supersoft' X-ray source (Greiner \& Teeseling 1998). 
Spectral fitting confirms that V~Sge's X-ray properties during
its soft X-ray state may be similar to those of supersoft X-ray
binaries, although a much lower luminosity cannot be excluded.

The model that has been suggested for \rxj0513\
cannot explain the observational data of V~Sge. First, the optical
brightness changes of V~Sge are very rapid: both the faint-/bright-state
transitions as well as the succession of different faint states may occur 
on timescales of \lax 1\,day (compared to the smooth decline
of several days in \rxj0513). Such very rapid changes are only possible
if the white dwarf envelope expands and contracts on the Kelvin-Helmholtz
timescale and the mass of the expanding envelope is rather small 
($M_{\rm env}\sim 10^{-9}$ \msun). Such a small envelope mass
is difficult to accept for a white dwarf with stable shell burning
(e.g. Prialnik \& Kovetz 1995). Second, the expected
optical eclipse would become deeper when the system becomes brighter,
opposite to what has been observed (Patterson \etal\ 1998).

It is possible to explain
the different optical/X-ray states of V~Sge by a variable amount of
extended uneclipsed matter,
which during the optical bright states contributes significantly
to the optical flux and completely absorbes the soft X-ray component
(Greiner \& Teeseling 1998).
A simple wind model for the recently observed radio flux density of V~Sge
implies a mass-loss rate of the order of 10$^{-6}$\,\msun/yr
(Lockley \etal\ 1997). With their (assumed) terminal velocity of 1500 km/s 
this wind zone is completely opaque for X-rays up to 0.7\,keV, even if the
wind is assumed to be circumbinary instead of arising from one component.
Since the radio measurement has been obtained during optical high state,
it supports the above described scenario.

\section{Discussion and Conclusion}

As shown above, both strategies to search for 
new SSS were successful. This is very promising, and opens up the way
to use optical observation strategies besides X-ray observations to 
identify new SSS.
The surprising part of this development is that both,
\v7\ and V Sge, are short-period CVs, and if true, their transient supersoft
source nature would establish a direct link between classical CVs and canonical
SSS which typically have orbital periods inthe 0.5--3 days range.

For both sources, V751 Cyg and V Sge, the estimates for their X-ray luminosity
(under reasonable assumptions on $N_{\rm H}$) are $\sim$ 
10$^{36}...10^{37}$ erg/s. This is at the lower end of the stable 
burning region. Two issues are relevant in this respect: 
(1) As for most CVs, the available evidence for VY Scl stars suggests that the
white dwarfs have a low mass.
At these low masses the accretion rate necessary for steady-state burning
(consistent with Fujimoto 1982) is
1--3$\times$10$^{-8}$ \msun/yr (Sion \& Starrfield 1994, Cassisi 1998), and
correspondingly the luminosity is lower than the canonical values for SSS.
(2) Even at rates below the steady-state burning level there could be a 
range where the hydrogen flashes are rather mild. A recent study of this
parameter space has shown that the luminosity between the flashes remains
at the surprisingly high value of $\sim$10\%--30\% of the burning luminosity 
(Rappaport 1999).

\noindent {\it Acknowledgement:}
I'm grateful to the organizers of this symposium for the exciting
atmosphere which stimulated many fruitful discussions.
JG is supported by the German Bun\-des\-mi\-ni\-sterium f\"ur Bildung,
Wissenschaft, Forschung und Technologie
(BMBF/DLR) under contract No. 50\,OR\,96\,09\,8.


\vspace{-0.1cm}
\begin{thebibliography}{Davidson \& Humphreys, 1997}

\vspace{-0.1cm}
\bibitem[]{brb95} Beuermann K., Reinsch K., Barwig H., \etal,
   1995, A\&A 294, L1

\bibitem[JJ]{b80} Bond H.E., 1980, 5th Ann. Workshop on Cataclysmic
   Variables and Related Objects (Austin, University of Texas)

\bibitem[]{cit98} Cassisi S., Iben I.Jr., Tornambe A., 1998, ApJ 496, 376


\bibitem[]{dr94} DiStefano R., Rappaport S., 1994, ApJ 437, 733

\bibitem[]{fuji82} Fujimoto M.Y., 1982, ApJ 257, 767

\bibitem[]{ghk91} Greiner J., Hasinger G., Kahabka P. 1991, A\&A 246, L17

\bibitem[]{gr95} Greiner J., 1995, Abano-Padova Conf. on Cataclysmic Variables,
eds. A. Bianchini \etal, 
ASSL 205, 443

\bibitem[]{gt98} Greiner J., van Teeseling A., 1998, A\&A 339, L21

\bibitem[]{gtd99} Greiner J., Tovmassian G.H., Di Stefano R., \etal,
  1999, A\&A 343, 183

\bibitem[]{kh97} Kahabka P., van den Heuvel E., 1997, ARA\&A 35, 69

\bibitem[]{king97} King A.R., 1997, MNRAS 288, L16

\bibitem[]{kl98} Knigge C., Livio M., 1998, MNRAS 297, 1079

\bibitem[]{lew97} Lockley J.J., Eyres S.P.S., Wood J.E., 1997, MNRAS 287, L14

\bibitem[]{mhp94} Motch C., Hasinger G., Pietsch W., 1994, A\&A 284, 827

\bibitem[]{pmb93} Pakull M.W., Motch C., Bianchi L., \etal,
              1993, A\&A 278, L39

\bibitem[]{pksts98} Patterson\,J., Kemp\,J., Shambrook\,A. \etal,
   1998, PASP 110, 380

\bibitem[]{pres98} Prestwich A.H., Silverman A., McDowell J., Callanan P., 
 Snowden S., 1999 (in prep.)

\bibitem[]{pk95} Prialnik D., Kovetz A., 1995, ApJ 445, 789

\bibitem[]{rap99} Rappaport S., 1999, these proceedings

\bibitem[]{rtba96} Reinsch K., van Teeseling A., Beuermann K., Abbott T.M.C., 
     1996, A\&A 309, L11

\bibitem[]{rhp97} Robertson J.W., Honeycutt R.K., Pier J.R., 1997, AJ 113, 787

\bibitem[]{sht93} Schaeidt S., Hasinger G., Tr\"{u}mper J., 1993, A\&A 270, L9

\bibitem[]{ss94} Sion E.M., Starrfield S.G., 1994, ApJ 421, 261

\bibitem[]{slc96} Southwell K.A., Livio M., Charles P.A., O'Donoghue D., 
   Sutherland W.J., 1996, ApJ 470, 1065

\bibitem[]{sd98} Steiner J.E., Diaz M.P., 1998, PASP 110, 276

\bibitem[]{} Tr\"{u}mper J., Hasinger G., Aschenbach B., \etal, 1991, 
    Nat 349, 579

\bibitem{vdhbnr92} van den Heuvel E.P.J., Bhattacharya D., Nomoto K., 
    Rappaport S.A.,  1992, A\&A 262, 97

\bibitem{trh97} van Teeseling A., Reinsch K., Hessman F.V., Beuermann K.,
   1997,  A\&A, L41

\bibitem[]{w95} Warner B., 1995, Cataclysmic Variable Stars, Cambridge 
   Astrophys. Ser. 28, Cambridge Univ. Press


\end{thebibliography}
\end{document}